\documentclass[11pt]{article}

\usepackage{fullpage}

\usepackage{algorithm}
\usepackage{algpseudocode}

\usepackage{booktabs} 

\usepackage{cite}

\usepackage{latexsym}
\usepackage{url}
\usepackage{listings}
\usepackage{xspace}
\usepackage{color}

\usepackage{graphicx}

\usepackage{amsmath}
\usepackage{amssymb}
\usepackage{amsthm}
\usepackage[binary-units]{siunitx}

\newcommand{\ceiling}[1]{\lceil #1\rceil}
\newcommand{\floor}[1]{\lfloor #1\rfloor}

\newcommand{\skips}[1]{\mathtt{skip[}#1\mathtt{]}}

\newcommand{\ablock}{\mathtt{block}}
\newcommand{\sendblock}[1]{\mathtt{sendblock[}#1\mathtt{]}}
\newcommand{\recvblock}[1]{\mathtt{recvblock[}#1\mathtt{]}}

\newcommand{\nextlink}[1]{\mathtt{next[}#1\mathtt{]}}
\newcommand{\prevlink}[1]{\mathtt{prev[}#1\mathtt{]}}

\newcommand{\mpibcast}{\textsf{MPI\_Bcast}\xspace}
\newcommand{\mpiallgatherv}{\textsf{MPI\_Allgatherv}\xspace}

\newcommand{\mpiint}{\textsf{MPI\_INT}\xspace}

\newcommand{\bidirec}[2]{\textsf{Send}(#1)\parallel\textsf{Recv}(#2)\xspace}

\newtheorem{theorem}{Theorem}
\newtheorem{proposition}{Proposition}
\newtheorem{lemma}{Lemma}
\newtheorem{observation}{Observation}

\newcommand{\gcc}{\texttt{gcc~10.2.1-6}\xspace}
\newcommand{\hydraopenmpi}{OpenMPI~4.1.4\xspace}

\DeclareSIUnit{\Gbps}{\giga\bit/\s}
\DeclareSIUnit{\microsecond}{\SIUnitSymbolMicro s}

\title{Round-optimal $n$-Block Broadcast Schedules in Logarithmic Time}
\author{Jesper Larsson Tr\"aff\\
  TU Wien\\
  Faculty of Informatics\\
  Institute of Computer Engineering, Research Group
  Parallel Computing 191-4\\
  Treitlstrasse 3, 5th Floor, 1040 Vienna, Austria}
\date{March-April 2023, December 2023}

\begin{document}
\maketitle

\begin{abstract}
  We give optimally fast $O(\log p)$ time (per processor) algorithms
  for computing round-optimal broadcast schedules for message-passing
  parallel computing systems.  This affirmatively answers the
  questions posed in Tr\"aff (2022).  The problem is to broadcast $n$
  indivisible blocks of data from a given root processor to all other
  processors in a (subgraph of a) fully connected network of $p$
  processors with fully bidirectional, one-ported communication
  capabilities. In this model, $n-1+\ceiling{\log_2 p}$ communication
  rounds are required. Our new algorithms compute for each processor
  in the network receive and send schedules each of size
  $\ceiling{\log_2 p}$ that determine uniquely in $O(1)$ time for each
  communication round the new block that the processor will receive,
  and the already received block it has to send. Schedule computations
  are done independently per processor without communication. The
  broadcast communication subgraph is the same, easily computable,
  directed, $\ceiling{\log_2 p}$-regular circulant graph used in
  Tr\"aff (2022) and elsewhere. We show how the schedule computations
  can be done in optimal time and space of $O(\log p)$, improving
  significantly over previous results of $O(p\log^2 p)$ and $O(\log^3
  p)$. The schedule computation and broadcast algorithms are simple to
  implement, but correctness and complexity are not obvious. All
  algorithms have been implemented, compared to previous algorithms,
  and briefly evaluated on a small $36\times 32$ processor-core
  cluster.
\end{abstract}

\section{Introduction}

We again consider the theoretically and practically immensely
important broadcasting problem for (subgraphs of) fully connected,
one-ported message-passing systems.

The broadcasting problem considered here is the following. In a
distributed memory system with $p$ processors, a designated root
processor has $n$ indivisible blocks of data that has to be
communicated to all other processors in the system. Each processor can
in a communication operation send an already known block to some other
processor and at the same time receive a(n unknown, new) block from
some other processor. Blocks can be sent and received in unit time,
where the time unit depends on the size of the blocks which are
assumed to all have (roughly) the same size. All processors can
communicate simultaneously.  Since communication of blocks takes the
same time, the complexity of an algorithm for solving the broadcast
problem can be stated in terms of the number of communication rounds
in which some or all processors are active that are required for the
last processor to have received all $n$ blocks from the root.  In this
fully-connected, one-ported, fully (send-receive) bidirectional $p$
processor system~\cite{BarNoyKipnis94,BarNoyKipnisSchieber00}, any
broadcast algorithm requires $n-1+\ceiling{\log_2 p}$ communication
rounds.  This follows from the observation that broadcasting a single
block requires $\ceiling{\log_2 p}$ communication rounds in any
one-ported system (since the number of processors that know the block
can at most double in a communication round). The last block can be
sent from the root after $n-1$ communication rounds and
$\ceiling{\log_2 p}$ final communication rounds are required for this
block to reach all other processors. A number of algorithms reach this
optimal number of communication rounds with different communication
patterns in a fully connected
network~\cite{BarNoyKipnisSchieber00,Jia09,KwonChwa95,Traff08:optibcast}.

The optimal communication round algorithm given
in~\cite{Traff08:optibcast} was used to implement the \mpibcast
operation for the Message-Passing Interface (MPI)~\cite{MPI-4.0}. Thus
a concrete, implementable solution was given, unfortunately with a
much too high schedule computation cost of $O(p\log^2 p)$ sequential
steps which could be amortized through careful
precomputation~\cite{Traff06:mpisxcoll}. An advantage of the algorithm
compared to other solutions to the broadcast
problem~\cite{BarNoyKipnisSchieber00,Jia09,KwonChwa95} is its simple,
$\ceiling{\log_2 p}$-regular circulant graph communication pattern
where all processors throughout the broadcasting operation operate
symmetrically (which has advantages for many different collective operations, see~\cite{Bruck97,BruckHo93,BarNoyKipnisSchieber93,Traff23:circulant}.
This makes it possible to use the algorithm for the
all-to-all broadcast problem and implement the difficult, irregular
\mpiallgatherv collective more efficiently as was shown recently
in~\cite{Traff22:bcastinprogress,Traff22:bcast}. In these papers, a
substantial improvement of the schedule computation cost was given,
from super-linear to $O(\log^3 p)$ time steps, thus presenting a
practically much more relevant algorithm. However, no adequate
correctness proof was presented. The other, major challenge posed in
these papers was to get the schedule computation down to $O(\log p)$
time steps. This is optimal, since at least $\ceiling{\log_2 p}$
communication rounds are required independently of $n$ in which each
processor sends and receives different blocks.

In this report, we prove the conjecture that correct and round optimal
send and receive schedules can be computed in $O(\log p)$ operations
per processor (without any communication) by stating and analyzing the
corresponding algorithms. The new algorithms use the same circulant
graph communication pattern and give rise to the same schedules as
those constructed by the previous algorithms of Tr\"aff et
al.~\cite{Traff08:optibcast,Traff22:bcastinprogress,Traff22:bcastba,Traff22:bcast}.
They are readily implementable, and of great practical relevance.

\section{Algorithms}

Assume that $n$ indivisible blocks of data have to be distributed,
either from a single, designated root processor, or from all
processors, to all other processors in a $p$-processor system with
processors $r,0\leq r<p$ that each communicate with certain other
processors by simultaneously sending and receiving data blocks.

We first show how broadcast from a designated root, $r=0$ (without
loss of generality) and all-to-all broadcast for any number of
processors $p$ can be done with regular, symmetric communication
patterns and explicit send and receive schedules that determine for
each communication operation by each processor which block is received
and which block is sent. We use these algorithms to formulate the
correctness conditions on possible send and receive schedules.

The communication pattern is then described concretely. Based on this
we present the two separate, explicit algorithms for computing
receive and send schedules that fulfill the correctness conditions. As
will be shown, these computations can be done fast in $O(\log p)$ time
per processor, independently of all other processors and with no
communication.

In all of the following, we let $p$ denote the number of processors,
and take $q=\ceiling{\log_2 p}$.

\subsection{Broadcast and all-to-all broadcast using schedules}
\label{sec:broadcasts}

\begin{algorithm}
  \caption{The $n$-block broadcast algorithm for processor $r,0\leq
    r<p$ of data blocks in array \texttt{buffer}. Round $x$ numbers
    the first round where actual communication takes place. Blocks
    smaller than $0$ are neither sent nor received, and for blocks
    larger than $n-1$, block $n-1$ is sent and received instead. Also
    blocks to the root processor are not sent. This is assumed to be
    taken care of by the bidirectional $\bidirec{}{}$ communication
    operations.}
  \label{alg:broadcast}
  \begin{algorithmic}
    \State$\Call{recvschedule}{r,\recvblock{}}$
    \State$\Call{sendschedule}{r,\sendblock{}}$
    \\
    \State $x\gets (q-(n-1+q)\bmod q)\bmod q$ \Comment Number of virtual rounds
    \State $i\gets 0$
    \While{$i<x$} \Comment Adjust schedule, $x$ virtual rounds already done
    \State$\mathtt{recvblock}[i]\gets\recvblock{i}-x+q$
    \State$\mathtt{sendblock}[i]\gets\sendblock{i}-x+q$
    \State $i\gets i+1$
    \EndWhile
    \While{$i<q$}
    \State$\mathtt{recvblock}[i]\gets\recvblock{i}-x$
    \State$\mathtt{sendblock}[i]\gets\sendblock{i}-x$
    \State $i\gets i+1$
    \EndWhile
    \State $i\gets x$
    \While{$i<n+q-1+x$}
    \State $k\gets i\bmod q$
    \State $t^k\gets (r+\skips{k})\bmod p$
    \Comment to- and from-processors
    \State $f^k\gets (r-\skips{k}+p)\bmod p$
    \\
    \State $\bidirec{\mathtt{buffer}[\sendblock{k}],t^k}{\mathtt{buffer}[\recvblock{k}],f^k}$
    \\
    \State $\sendblock{k}\gets \sendblock{k}+q$
    \State $\recvblock{k}\gets \recvblock{k}+q$
    \State $i\gets i+1$
    \EndWhile
\end{algorithmic}
\end{algorithm}

Generic algorithms for $n$ block broadcast and all-to-all broadcast
communication operations are shown as Algorithm~\ref{alg:broadcast}
and Algorithm~\ref{alg:irregallgather} (and were also explained
in~\cite{Traff22:bcastinprogress,Traff22:bcast}). Both algorithms are
symmetric in the sense that all processes follow the same regular
graph communication pattern and do the same communication operations
in each round. For the rooted, asymmetric broadcast operation, this is
perhaps surprising.

The idea of the algorithms is as follows. The processors communicate
in rounds, starting from some round $x$ (to be explained shortly) and
ending at $n-1+q+x$ for a total of the required $n-1+q$ communication
rounds. For round $i,x\leq i<n-1+q+x$, we take $k=i\bmod q$, such that
always $0\leq k<q$. In round $i$, each processor $r, 0\leq r<p$
simultaneously sends a block to a \emph{to-processor}
$t^k=(r+\skips{k})\bmod p$ and receives a different block from a
\emph{from-processor} $f^k =(r-\skips{k}+p)\bmod p$, determined by a
skip per round $\skips{k}, 0\leq k<q$.  The blocks that are sent and
received are numbered consecutively from $0$ to $n-1$ and stored in a
\texttt{buffer} array indexed by the block number. The block that a
processor sends in round $i$ is determined by a send schedule array
$\sendblock{k}$ and likewise the block that a processor will receive
in round $i$ by a receive schedule array $\recvblock{k}$. Since the
blocks are thus fully determinate, no block indices or other meta-data
information is ever communicated by the algorithms. The $\sendblock{}$
and $\recvblock{}$ arrays are computed such that blocks are
effectively sent from root processor $r=0$ that initially has all $n$
blocks, and such that all $n$ blocks are received and sent further on
by all the other processors. The starting round $x$ is chosen such
that $(n-1+q+x)\bmod q=0$ and after this last round which is a
multiple of $q$, all processors will have received all $n$ blocks. The
assumption that processor $r=0$ is the root can be made without loss
of generality. Should some other processor $r'$ be root, the
processors are simply renumbered by subtracting $r'$ (modulo $p$) from
the processor indices.

The broadcast algorithm is shown as Algorithm~\ref{alg:broadcast}.
Not shown in Algorithm~\ref{alg:broadcast} is that no block is ever
sent back to the root which already has all the blocks in the first
place (so no send operation if $t^k=0$), and that non-existent,
negatively indexed blocks are never sent nor received (if
$\sendblock{k}<0$ or $\recvblock{k}<0$ for some $k$, the corresponding
send and receive communication is simply ignored). For block indices
larger than the last block $n-1$, block $n-1$ is instead sent and
received. These cases are assumed to be handled by the concurrent
send- and receive operations as indicated by $\bidirec{}{}$. The
receive and send block schedules $\recvblock{}$ and $\sendblock{}$ are
computed by the calls to \textsc{recvschedule()} and
\textsc{sendschedule()} functions to be derived in
Section~\ref{sec:recvschedule} and Section~\ref{sec:sendschedule}.

For the algorithm to be correct (in the sense of broadcasting all
blocks from processor $r=0$ to all other processors), the following
conditions must hold:
\begin{enumerate}
  \item
    The block that is received in round $i$ with $k=i\bmod q$ by some
    processor $r$ must be the block that is sent by the from-processor
    $f_r^k$, $\recvblock{k}_r = \sendblock{k}_{f_r^k}$. Equivalently,
  \item
    the block that processor $r$ sends in round $i$ with $k=i\bmod q$
    must be the block that the to-processor $t_r^k$ will receive,
    $\sendblock{k}_r = \recvblock{k}_{t_r^k}$.
  \item
    \label{cor:qblocks}
    Over any $q$ successive rounds, each processor must receive $q$
    different blocks. More concretely,
    $\bigcup_{k=0}^{q-1}\recvblock{k}
    =(\{-1,-2,\ldots,-q\}\setminus\{b-q\})\cup\{b\}$ where $b, 0\leq
    b<q$ is the first actual, non-negative block received by the
    processor in one of the first $q$ rounds. This block $b$ is called
    the \emph{baseblock} for processor $r$.
  \item
    \label{cor:recvbefore}
    The block that a processor sends in round $i$ with $k=i\bmod q$
    must be a block that has been received in some previous round, so
    either $\sendblock{k}=\recvblock{j}$ for some $j, 0\leq j<k$, or
    $\sendblock{k}=b-q$ where $b\geq 0$ is the \emph{baseblock} of
    the processor.
\end{enumerate}
The last correctness condition implies that $\sendblock{0}=b-q$ for
each processor. With receive and send schedules fulfilling these four
conditions, it is easy to see that the broadcast algorithm in
Algorithm~\ref{alg:broadcast} correctly broadcasts the $n$ blocks over the
$p$ processors.

\begin{theorem}
  Let $K,K>0$ be a number of communication phases each consisting of
  $q$ communication rounds for a total of $Kq$ rounds. Assume that in
  each round $i, 0\leq i<Kq$, each processor $r,0\leq r<p$ receives a
  block $\recvblock{i\bmod q}+\floor{i/q}q$
  and sends a block
  $\sendblock{i\bmod q}+\floor{i/q}q$
  (provided these blocks are non-negative). By the end of the $Kq$
  rounds, processor $r$ will have received all blocks
  $\{0,1,\ldots,(K-1)q-1\}\cup \{b+(K-1)q\}$ where $b$ is the first
  (non-negative) block received by processor $r$.
\end{theorem}
\begin{proof}
  The proof is by induction on the number of phases. For $K=1$, there
  are $q$ rounds $i=0,1,\ldots,q-1$ over which each processor will
  receive its non-negative baseblock $b$; all other receive blocks are
  negative (Correctness Condition~\eqref{cor:qblocks}). For $K>1$, in
  the last phase $K-1$, each processor will receive the blocks
  $(\{(K-2)q,(K-2)q+1,\ldots,(K-2)q+q-1\}\setminus\{b+(K-2)q\})\cup\{b+(K-1)q\}$
  since the set $\bigcup_{k=0}^{q-1}\recvblock{k}$ contains $q$
  different block indices, one of which is positive. The block
  $b+(K-2)q$ has been received in phase $K-2$ by the induction
  hypothesis, in its place block $b+(K-1)q$ is received. Therefore, at
  the end of phase $K-1$ using the induction hypothesis, blocks
  \begin{displaymath}
    \{0,1,\ldots,(K-2)q-1\}\cup \{(K-2)q,(K-2)q+1,\ldots,(K-2)q+q-1\}
    = \{0,1,\ldots,(K-1)q-1\}
  \end{displaymath}
  plus the block $b+(K-1)q$ have been received, as claimed. By
  Correctness Condition~\eqref{cor:recvbefore}, no block is sent that has
  not been received in a previous round or phase.
\end{proof}
In order to broadcast a given number of blocks $n$ in the optimal
number of rounds $n-1+q$, we use the smallest number of phases $K$
such that $Kq\geq n-1+q$, and introduce a number of dummy blocks
$x=Kq-(n-1+q)$ that do not have to be broadcast. In the $K$ phases,
all processors will have received $n+x-1$ blocks $0,1,\ldots,n+x-2$
plus one larger block. We perform $x$ initial, virtual rounds with no
communication to handle the $x$ dummy blocks, and broadcast the actul,
non-negative blocks in the following $n-1+q$ rounds. This is handled
by simply subtracting $x$ from all computed block indices; negative
blocks are neither received nor sent. Blocks with index larger than
$n-1$ in the last phase are capped to $n-1$.

The symmetric communication pattern where each processor (node) $r$
has incoming (receive) edges $(f_r^k,r)$ and outgoing (send) edges
$(r,t^k_r)$ is a \emph{circulant graph} with skips (jumps)
$\skips{k},k=0,1,\ldots, q-1$. The $q$ skips for the circulant graph
are explained and computed in Section~\ref{sec:circulant}.

\begin{algorithm}
  \caption{The $n$-block all-to-all broadcast algorithm for processor
    $r,0\leq r<p$ for data in the arrays $\mathtt{buffers}[j],0\leq
    j<p$. The count $x$ is the number of empty first rounds.  Blocks
    smaller than $0$ are neither sent nor received, and for blocks
    larger than $n-1$, block $n-1$ is sent and received instead.}
  \label{alg:irregallgather}
  \begin{algorithmic}
    \For{$j=0,1,\ldots,p-1$}
    \State $r'\gets (r-j+p)\bmod p$
    \State$\Call{recvschedule}{r',\mathtt{recvblocks}[j][]}$
    \EndFor
    \For{$j=0,1,\ldots,p-1$}
    \For{$k=0,1,\ldots,q-1$}
    \State $f^k\gets (j-\skips{k}+p)\bmod p$
    \State$\mathtt{sendblocks}[j][k]\gets\mathtt{recvblocks}[f^k][k]$
    \EndFor
    \EndFor
    \\
    \State $x\gets (q-(n-1+q)\bmod q)\bmod q$
    \Comment Number of virtual rounds
    \For{$j=0,1,\ldots, p-1$}
    \State $i\gets 0$
    \While{$i<x$} \Comment Adjust schedules, $x$ virtual rounds already done
    \State $\mathtt{recvblocks}[j][i]\gets\mathtt{recvblocks}[j][i]-x+q$
    \State $\mathtt{sendblocks}[j][i]\gets\mathtt{sendblocks}[j][i]-x+q$
    \State $i\gets i+1$
    \EndWhile
    \While{$i<q$}
    \State $\mathtt{recvblocks}[j][i]\gets\mathtt{recvblocks}[j][i]-x$
    \State $\mathtt{sendblocks}[j][i]\gets\mathtt{sendblocks}[j][i]-x$
    \State $i\gets i+1$
    \EndWhile
    \EndFor
    \State $i\gets x$
    \While{$i<n+q-1+x$}
    \State $k\gets i\bmod q$
    \State $t^k, f^k\gets (r+\skips{k})\bmod p,(r-\skips{k}+p)\bmod p$ \Comment to- and from-processors
    \\
    \State $j'\gets 0$
    \For{$j=0,1,\ldots,p-1$} \Comment Pack
    \If{$j\neq t^k$} $\mathtt{tempin}[j'],j'\gets\mathtt{buffers}[j][\mathtt{sendblocks}[j][k]],j'+1$
    \EndIf
    \State $\mathtt{sendblocks}[j][k]\gets \mathtt{sendblocks}[j][k]+q$ \EndFor

    \State $\bidirec{\mathtt{tempin},t^k}{\mathtt{tempout},f^k}$

    \State $j'\gets 0$ \For{$j=0,1,\ldots,p-1$} \Comment Unpack
    \If{$j\neq r$}
    $\mathtt{buffers}[j][\mathtt{recvblocks}[j][k]],j'\gets\mathtt{tempout}[j'],j'+1$
    \EndIf
    \State $\mathtt{recvblocks}[j][k]\gets \mathtt{recvblocks}[j][k]+q$
    \EndFor
    \State $i\gets i+1$
    \EndWhile
\end{algorithmic}
\end{algorithm}

The explicit send and receive schedules can be used for all-to-all
broadcast as shown as Algorithm~\ref{alg:irregallgather}. In this
problem, each processor has $n$ blocks of data to be broadcast to all
other processors. The $n$ blocks for each processor $r$ are as for the
broadcast algorithm assumed to be roughly of the same size, but blocks
from different processors may be of different size as long as the same
number of $n$ blocks are to be broadcast from each processor. The
algorithm can therefore handle the irregular case where different
processors have different amounts of data to be broadcast as long as
each divides its data into $n$ roughly equal-sized blocks.  Due to the
fully symmetric, circulant graph communication pattern, this can be
done by doing the $p$ broadcasts for all $p$ processors $r, 0\leq r<p$
simultaneously, in each communication step combining blocks for all
processors into a single message.  The blocks for processor $j$ are
assumed to be stored in the buffer array $\mathtt{buffers}[j][]$
indexed by block numbers from $0$ to $n-1$.  Initially, processor $r$
contributes its $n$ blocks from $\mathtt{buffers}[r][]$. The task is
to fill all other blocks $\mathtt{buffers}[j][]$ for $j\neq r$.  Each
processor $r$ computes a receive schedule $\mathtt{recvblocks}[j]$ for
each other processor as root processor $j,0\leq j<p$ which is the
receive schedule for $r'=(r-j+p)\bmod p$.  Note that this indexing is
slightly different from the algorithm as stated
in~\cite{Traff22:bcastinprogress,Traff22:bcast}.  Before each
communication operation, blocks for all processors $j,0\leq j<p$ are
packed consecutively into a temporary buffer \texttt{tempin}, except
the block for the to-processor $t^k$ for the communication round. This
processor is the root for that block, and already has the
corresponding block. After communication, blocks from all processors
are unpacked from the temporary buffer \texttt{tempout} into the
$\mathtt{buffers}[j][]$ arrays for all $j,0\leq j<p$ except for $j=r$:
A processor does not receive blocks that it already has. As in the
broadcast algorithm in Algorithm~\ref{alg:broadcast}, it is assumed
that the packing and unpacking will not pack for negative block
indices, and that indices larger than $n-1$ are taken as $n-1$. Also
packing and unpacking blocks for processors not contributing any data
(as can be the case for highly irregular applications of all-to-all
broadcast) shall be entirely skipped (not shown in
Algorithm~\ref{alg:irregallgather}), so that the total time spent in
packing and unpacking per processor over all communication rounds is
bounded by the total size of all $\mathtt{buffers}[j][], j\neq t^k$
and $\mathtt{buffers}[j][], j\neq r$.

\subsection{The communication pattern}
\label{sec:circulant}

\begin{algorithm}
  \caption{Computing the skips (jumps) for a $p$-processor circulant
    graph ($q=\ceiling{\log_2 p}$).}
  \label{alg:circulants}
\begin{algorithmic}
  \State $k\gets q$ \State $\skips{k}\gets p$ \While{$k>0$} \State
  $k\gets k-1$ \State $\skips{k}\gets \skips{k+1}-\skips{k+1}/2$
  \EndWhile
\end{algorithmic}
\end{algorithm}

The skips for the circulant graph communication pattern are computed
by repeated halving of $p$ as shown as Algorithm~\ref{alg:circulants}.
For convenience, we take $\skips{q}=p$. The algorithm iterates
downwards from $k=q-1$, in each iteration dividing the previous
$\skips{k+1}$ by two and rounding up, here expressed by integer
floor-division. It can easily be seen (by induction) that
$q=\ceiling{\log_2 p}$ halving steps are necessary and sufficient to
get $\skips{0}=1$ (the induction hypothesis being that for
$2^{q-1}<p\leq 2^q$, $q$ halving steps are required). We make a number
of observations that are necessary for developing the receive and send
schedules in the following.

\begin{observation}
  \label{obs:skips}
  For each $k, 0\leq k<q$ it holds that
  $\skips{k}+\skips{k}\geq\skips{k+1}$
\end{observation}
This follows directly from the halving scheme of
Algorithm~\ref{alg:circulants}. If $\skips{k+1}$ is even, the halving
is exact and $\skips{k}+\skips{k}=\skips{k+1}$, and otherwise
$\skips{k}+\skips{k}=\skips{k+1}+1>\skips{k+1}$.

\begin{observation}
  \label{obs:tunnel}
  For any $p$, there are at most two $k, k>1$ such that
  $\skips{k-2}+\skips{k-1}=\skips{k}$.
\end{observation}
For $\skips{2}=3$, Algorithm~\ref{alg:circulants} gives $\skips{1}=2$
and $\skips{0}=1$ for which the observation holds. For $\skips{3}=5$,
Algorithm~\ref{alg:circulants} gives $\skips{2}=3$ and $\skips{1}=2$
for which the observation holds. Any $p$ for which $\skips{2}=3$ or
$\skips{3}=5$ will have this property, and none other. We see that for
all $p$, $\skips{0}=1$ and $\skips{1}=2$, and that $\skips{2}\geq 3$,
$\skips{3}\geq 5$ and $\skips{4}\geq 9$, and therefore
$\skips{k-2}+\skips{k-1}<\skips{k}$ for $k>3$ for all $p$.

\begin{observation}
  \label{obs:cut}
  For some $p$ and $k>0$, there is an $r, r<\skips{k}$ with
  $r+\skips{k}=\skips{k+1}$.
\end{observation}
If $\skips{k+1}$ is odd, $r=\skips{k+1}-\skips{k}$ fulfills the
observation.

\begin{observation}
  \label{obs:chains}
  For each $k,0\leq k<q$ it holds that
  $1+\sum_{i=0}^{k-1}\skips{i}\geq \skips{k}$.  For each $k,0<k<q$, it
  holds that $\sum_{i=0}^{k-2}\skips{i}<\skips{k}$.
\end{observation}
The claims follow easily by induction with the previous observations
as induction bases. Namely, $1+\sum_{i=0}^{k-1}\skips{i}+\skips{k}\geq
\skips{k}+\skips{k}\geq\skips{k+1}$, and
$\sum_{i=0}^{k-2}\skips{i}+\skips{k-1}<\skips{k-1}+\skips{k}\leq\skips{k+1}$.

\begin{observation}
  \label{obs:admit}
  If $\skips{e}+\skips{k}<r$ for some $e, 0<e<q$ and $k, k<e$, then also
  $\skips{e-i}+\skips{k+i}<r$ for $i=0,1,\ldots,e-k$.
\end{observation}

\begin{lemma}
  \label{lem:canonical}
  For any $r,0\leq r <p$ there is a (possibly empty) sequence
  $[e_0,e_1,\ldots,e_{j-1}]$ of $j, j<q$, different skip indices such
  that $r=\sum_{i=0}^{j-1} \skips{e_i}$.
\end{lemma}
We call a (possibly empty) sequence $[e_0,e_1,\ldots,e_{j-1}]$ for
which $r=\sum_{i=0}^{j-1} \skips{e_i}$ and where
$e_0<e_1<\ldots<e_{j-1}$ a \emph{skip sequence} for $r$.

\begin{proof}
  The proof is by induction on $k$. When $r=0$ the claim holds for the
  empty sequence. Assuming the claim holds for any $r, 0\leq
  r<\skips{k}$, we show that it holds for $0\leq r<\skips{k+1}$. If
  already $0\leq r<\skips{k}$ the claim holds by assumption.  If
  $\skips{k}\leq r <\skips{k+1}$, then $0\leq
  r-\skips{k}<\skips{k+1}-\skips{k}\leq\skips{k}$ by
  Observation~\ref{obs:skips}. By the induction hypothesis, there is a
  sequence of different skips not including $\skips{k}$ and summing to
  $r-\skips{k}$, and $\skips{k}$ can be appended to this sequence to
  sum to $r$.
\end{proof}

The lemma indicates how to recursively compute in $O(q)$ steps a
specific, \emph{canonical skip sequence} for any $r, 0\leq r<p$. By
Observation~\ref{obs:tunnel} and Observation~\ref{obs:cut}, for some
$p$ there may be more than one skip sequence for some $r$; the
decomposition of $r$ into a sum of different skips is not unique for
all $p$ (actually, the decomposition is unique only when $p$ is a
power of $2$).  A canonical skip sequence will contain $\skips{k}$ and
not $\skips{k-2}$ and $\skips{k-1}$ if
$\skips{k-2}+\skips{k-1}=\skips{k}$ (Observation~\ref{obs:tunnel}),
and $\skips{k+1}$ instead of $\skips{k}$ if $r+\skips{k}=\skips{k+1}$
(Observation~\ref{obs:cut}).

A non-empty skip sequence $[e_0,e_1,\ldots,e_{j-1}]$ for $r$ defines a
path from processor $0$ to processor $r>0$ as follows. From $0$ to
$\skips{e_{0}}$ through edge $(0,\skips{e_{0}})$, from $\skips{e_{0}}$
to $(\skips{e_{0}}+\skips{e_{1}})\bmod p$ through edge
$(\skips{e_0},(\skips{e_0}+\skips{e_1})\bmod p)$ and so on.  The edges
on the path to $r$ are
$((\sum_{j=0}^{i-1}\skips{e_i})\bmod p),(\sum_{j=0}^{i}\skips{e_i})\bmod p)$
for $i=0,\ldots,j-1$. Note that the skips along the path strictly increase.
We will use the terms skip sequence and path interchangeably.

\begin{algorithm}
  \caption{Finding the baseblock for processor $r,0\leq r<p$.}
  \label{alg:baseblock}
\begin{algorithmic}[1]
  \Function{baseblock}{$r$}
  \State $k\gets q$
  \Repeat
  \State $k\gets k-1$
  \If{$\skips{k}=r$} \Return{$k$}
  \ElsIf{$\skips{k}<r$} $r\gets r-\skips{k}$
  \EndIf
  \Until{$k=0$}
  \State \Return{$q$} \Comment Only processor $r=0$ will return $q$ as baseblock
  \EndFunction
\end{algorithmic}
\end{algorithm}

The canonical skip sequence for an $r,0\leq r<p$ is implicitly
computed iteratively by the \textsc{baseblock()} function of
Algorithm~\ref{alg:baseblock} which explicitly returns the first
(smallest) skip index in the canonical skip sequence.  This index is
called the \emph{baseblock} for $r$ and is of vital importance for the
broadcast schedules. For convenience, we define $q$ to be the
baseblock of $r=0$ for which the skip sequence is otherwise empty; for
other $r>0$ the baseblock $b$ satisfies $0\leq b<q$ and is a legal
skip index. 

If we let the root processor $r=0$ send out blocks one after the other
(that is, $\sendblock{k}=k, k=0,1,\ldots,q-1$) on the edges
$(0,\skips{k})$, the canonical skip sequence for any other $r, r>0$
gives a path through which the baseblock $b$ for $r$ can be sent from
the root to processor $r$. Processor $r$ will receive its baseblock in
communication round $e$ where $e$ is the last, largest skip index in
the skip sequence for $r$, therefore $\recvblock{e}_r=b$.  In the
broadcast schedules that will be used in Algorithm~\ref{alg:broadcast}
and Algorithm~\ref{alg:irregallgather}, the baseblock $b$ for each
processor $r$ is therefore the first actual, non-negatively indexed block
that the processor receives; in each of the following rounds, $r$ will
be be receiving new blocks different from the baseblock.


\subsection{The receive schedule}
\label{sec:recvschedule}

\begin{algorithm}
  \caption{Computing receive blocks for processor $r, p\leq r<2p$ by
    depth-first search with removal of accepted blocks.}
  \label{alg:dfspath}
  \begin{algorithmic}[1]
    \Function{DFS-blocks}{$r,r',s,e,k,\recvblock{}$}
    \If{$r'\leq r-\skips{k+1}$} 
    \While{$e\neq -1$}
    \If{$r'+\skips{e}\leq r-\skips{k}$} \Comment Index $e$ admissible for $k$
    \State $k\gets \Call{DFS-blocks}{r,r'+\skips{e},s,e,k,\recvblock{}}$
    \State\Comment Even if $k$ has changed, admissibility $r'+\skips{e}\leq r-\skips{k}$ still holds
    \If{$r'\leq r-\skips{k+1}\wedge s>r'+\skips{e}$} \Comment Canonical path found
    \State $s\gets r'+\skips{e}$
    \State $\recvblock{k},k\gets e,k+1$ \Comment Accept $e$, next $k$
    \State $\nextlink{\prevlink{e}},\prevlink{\nextlink{e}}\gets \nextlink{e},\prevlink{e}$ \Comment Remove $e$ by unlinking 
    \EndIf
    \EndIf
    \State $e\gets \nextlink{e}$
    \EndWhile
    \EndIf
    \State \Return $k$
    \EndFunction
  \end{algorithmic}
\end{algorithm}

We now show how to compute the receive schedule $\recvblock{k},
k=0,\ldots q-1$ for any processor $r,0\leq r<p$ in $O(q)$
operations. More precisely, we compute for any given $r$ the $q$
blocks that $r$ will receive in the $q$ successive communication
rounds $k=0,1,\ldots,q-1$ when processor $0$ is the root
processor. The basis for the receive schedule computation is to find
$q$ paths from the root to $r$ in the form of canonical skip sequences
to intermediate processors $r', r'<r$. For $q$ skips, there are
obviously $2^q\geq p$ (but $<2p$) canonical skip sequences, so
exploring them all (for instance, by depth-first search) will give a
linear time (or worse) and not an $O(\log p)$ time algorithm.

Instead, a greedy search through the skip sequences and paths is done
by a special backtracking algorithm. The complexity of the computation
is reduced by removing the first (smallest) skip index, corresponding
to the baseblock for $r'$, each time a good canonical skip sequence to
some $r'<r$ has been found.  The first skip index of this $k$th
canonical sequence shall be taken as the block sent by the root that
eventually arrives at processor $r$ in the $k$th round,
$k=0,1,\ldots,q-1$. This will guarantee that there are indeed $q$
different blocks among $\recvblock{k}$ as required by Correctness
Condition~\eqref{cor:qblocks}.

More concretely, the backtrack search finds a canonical skip sequence
summing to the $r'$ with $r'\leq r-\skips{k}$ that is closest to
$r-\skips{k}$ using only skips that have not been removed from paths
closest to $r-\skips{j}$ for $j<k$ already found. The processor from
which $r$ will receive its $k$th block is indeed $r-\skips{k}$, and
therefore it must hold that $r'\leq r-\skips{k}$.  This depth-first
like search with removal is shown as Algorithm~\ref{alg:dfspath}. From
this $r'$, we will show that there is a canonical skip sequence to $r$
consisting of only some skips $j$ for $0\leq j<k$.

The \textsc{DFS-blocks()} function assumes that the (remaining) skip
indices are in a doubly linked list in decreasing order. Thus, for
skip index $e$, $\nextlink{e}$ is the next, smaller, remaining skip
index. This list is used to try the skips in decreasing order as in
Algorithm~\ref{alg:baseblock}. An index is removed by linking it out
of the doubly linked list and can easily be done in $O(1)$ time as
shown in Algorithm~\ref{alg:dfspath}. For each $k, k\geq 0$, the
function recursively and greedily searches for a largest $r'$ with
$r'\leq r-\skips{k}$ similarly to the baseblock computation in
Algorithm~\ref{alg:baseblock}. The last (smallest) skip index $e$ for
which this is the case will be taken as the $k$th receive block
$\recvblock{k}$ and removed from the list of skip indices. The
corresponding recursive call terminates and the algorithm backtracks
to find the receive block for $k+1$. The depth first search greedily
increases a found $r'$ by the largest remaining $\skips{e}$ for which
$r'+\skips{e}\leq r-\skips{k}$. In order to ensure that the canonical
path from $r'$ to $r$ consists of only skip indices $j<k$, $e$ is
accepted only if also $r'\leq r-\skips{k+1}$. This means that even
when $r'+\skips{e}$ is accepted as the processor closest to $r$, there
is still a path from $r'$ to $r$ via $\skips{k+1}$.  After the
recursive call, skip index $e$ is accepted if the path $r'+\skips{e}$
is not equal to the length of last found path closest to $r-\skips{k}$. This is
necessary to ensure that the path found is canonical; according to
Observation~\ref{obs:cut} and Observation~\ref{obs:tunnel} there may
be more than one path to $r'+\skips{e}$ and the canonical one has to
be chosen. The smallest skip index $e$ extending the path to $r'$
fulfilling the conditions is the receive block for round $k$ and is
stored in $\recvblock{k}$.

\begin{algorithm}
  \caption{Computing the receive schedule for processor $r,0\leq r<p$.}
  \label{alg:recvschedule}
  \begin{algorithmic}
    \Procedure{recvschedule}{$r,\recvblock{}$}
    \State\Comment Build doubly linked list to scan skips in decreasing order
    \For{$e=0,\ldots,q$}
    \State $\nextlink{e},\prevlink{e}\gets e-1, e+1$
    \EndFor
    \State $\prevlink{q}\gets -1$
    \State $\nextlink{-1},\prevlink{-1}\gets q,0$
    \\
    \State $b\gets\Call{baseblock}{r}$
    \State $\nextlink{\prevlink{b}},\prevlink{\nextlink{b}}\gets \nextlink{b},\prevlink{b}$ \Comment Remove baseblock index $b$ by unlinking
    \State\Call{DFS-blocks}{$p+r,0,p+p,q,0,\recvblock{}$}  \Comment Ensure $q$ blocks
    \For{$k=0,\ldots,q-1$}
    \If{$\recvblock{k}=q$} $\recvblock{k}\gets b$
    \Else\ $\recvblock{k}\gets\recvblock{k}-q$
    \EndIf
    \EndFor
    \EndProcedure
  \end{algorithmic}
\end{algorithm}

The \textsc{DFS-blocks()} algorithm is now used to compute the receive
schedule for a processor $r,0\leq r<p$ as shown in
Algorithm~\ref{alg:recvschedule}. In order to avoid problems with
$r-\skips{k}$ becoming negative, we instead compute the sequence of
closest $r'$ for (virtual) processor $p+r$.  The algorithm uses the
$q+1$ skips computed by Algorithm~\ref{alg:circulants} (including
$\skips{q}=p$) and searches for canonical paths to $p+r$. In order to
exclude the canonical path leading to $r$ itself with baseblock $b$
(as computed by Algorithm~\ref{alg:baseblock}), $b$ is removed from
the list of skip indices before calling \textsc{DFS-blocks}. For the initial
call to the recursive procedure, there is no previous path, so both $r'=0$ and
$s=0$. The search starts from the largest skip index $e$ with $k=0$.

Called this way, upon return the \textsc{DFS-blocks()} function
obviously returns $q$ different, positive skip indices in
$\recvblock{k}, 0\leq k<q$.  In Algorithm~\ref{alg:broadcast} and
Algorithm~\ref{alg:irregallgather}, in the first $q$ communication
rounds where $k=0,1,\ldots,q-1$, only the (positive) baseblocks will
be received by the processors, while all other blocks are blocks that
will be received in the next $q,q+1,\ldots,2q-1$ rounds. Therefore,
$q$ is subtracted from the block indices, except for baseblock $q$ in
some $\recvblock{k}$. This block corresponds to the round $k$ where
$p+0=\skips{q}$ is the processor closest to $r-\skips{k}$ (not using
the skip indices removed before round $k$), so this $k$ is the round
where $r$ will receive its baseblock from the root. For this $k$,
$\recvblock{k}$ is set to $b$.

\begin{proposition}
  \label{prop:recvtime}
  When called as $\textsc{DFS-blocks}(p+r,0,p+p,q,0,\recvblock{})$ with a
  list of skip indices in decreasing order that excludes the baseblock $b$ of
  $r$, Algorithm~\ref{alg:dfspath} computes $q$ different blocks
  $\{0,1,\ldots,q\}\setminus\{b\}$ in $\recvblock{}$ in $O(\log p)$
  operations.
\end{proposition}
\begin{proof}
  We first prove that a simplified version of
  Algorithm~\ref{alg:dfspath} fulfills the claim and performs $q$
  recursive calls. Each recursive call is with a new $r'+\skips{e}$
  that is closer to $r-\skips{k}$.  To this end, we say that a skip
  index $e$ is \emph{admissible for $k$} if $r'+\skips{e}\leq
  r-\skips{k}$, and for now ignore the further conditions $r'\leq
  r-\skips{k+1}$ and $s>r'+\skips{e}$ for accepting a skip index to
  the canonical skip sequence. These conditions are necessary to
  ensure that a computed skip sequence is indeed canonical as defined
  by Lemma~\ref{lem:canonical}.  In the simplified case, each $e$ is
  removed after the recursive call, and therefore never considered
  again (in the while loop of some previous recursive call).  Each $e$
  becomes admissible once for some $k$ (since the
  \textsc{DFS-blocks()} is called with processor $p+r$ and
  $\skips{k}\leq p$), therefore the number of recursive calls is $q$.
  This claim is justified in Lemma~\ref{lem:admissibility}.

  In the non-simplified version of the algorithm, throughout the
  recursive calls, $s$ is the sum of the skips on the most recently
  accepted path.  If an admissible skip index $e$ is not accepted
  because $s=r'+\skips{e}$ (which could be the case if
  $\skips{k-2}+\skips{k-1}=\skips{k}$, Observation~\ref{obs:tunnel},
  or $r'+\skips{k}=\skips{k+1}$, Observation~\ref{obs:cut}) another
  recursive call on $e$ is necessary on some other path. Indeed, $e$
  will be admissible in the parent recursive call. In
  the same way, a recursive call that is terminated because
  $r'>r-\skips{k+1}$ will have to be repeated. This can happen at most
  once for each skip index, since also here $e$ will 
  be admissible in the parent recursive call.

  Algorithm~\ref{alg:dfspath} therefore performs at most $2q$
  recursive calls which can be done $O(\log p)$ operations since there
  are at most $q$ additional \textbf{while}-loop iterations over
  all calls where skip indices are not admissible.
\end{proof}

In order to ensure that one recursive call per skip index $e$ suffices
(in the simplified version of Algorithm~\ref{alg:dfspath}, two in the
full version), it needs to be shown that if skip index $e$ is
admissible for $k$ before the recursive call, it has not been removed
and will still be admissible for a possibly larger $k'$ upon return.

\begin{lemma}
  \label{lem:admissibility}
  If skip index $e$ is admissible for $k$ and therefore leading to a
  recursive \textsc{DFS-blocks()} call on $r'+\skips{e}$, index $e$
  will be admissible for the $k'\geq k$ returned by the call. By the
  end of the \textbf{while}-loop of the call, all skip indices
  $e,\ldots,0$ will have been removed, and $k'=e+1$.
\end{lemma}

\begin{proof}
  The proof is by structural induction on the recursive calls to
  \textsc{DFS-blocks()}. Consider the first recursive call that does
  not cause a further recursive call (which costs $e-1$ unsuccessful
  admissibility checks in the loop of the call). By return from this
  call, $k$ is unchanged and so the skip index $e$ remains admissible
  since it still holds that $r'+\skips{e}\leq r-\skips{k}$. Therefore,
  index $e$ will be taken as $\recvblock{k}$ and $k$ incremented.
  Admissibility means that $\skips{e}+\skips{k}\leq r-r'$. By
  Observation~\ref{obs:admit}, all smaller $e$ in the remainder of the
  \textbf{while}-loop will also be admissible, if they cause no
  further recursive calls. However, if $e>3$, there will be a chain of
  recursive calls $r'+\sum^{e-1}_{i=0}\skips{i}<r-\skips{1}$ each of which
  will be admissible and cause removal of the skip indices $i$ in constant
  time. Thus, by the end of the \textbf{while}-loop, all skip indices
  $0,1,\ldots,e$ will have been removed, and $k'=e+1$.

  Consider now a recursive call on an admissible skip index $e$ that
  causes further recursive calls, and let $e', e>e'$ be the skip index
  of the first such call. Before the recursive call,
  $r'+\skips{e}+\skips{e'}\leq r-\skips{k}$. By the induction
  hypothesis, upon return from this call $e'$ will be admissible and
  removed and return with $k'=e'+1$. As $e'$ was admissible,
  $r'+\skips{e}+\skips{e'}\leq r-\skips{e'}$, and therefore
  $r'+\skips{e}\leq r-\skips{e'}-\skips{e'}\leq r-\skips{e'+1}$ since
  by Observation~\ref{obs:skips}
  $\skips{e'}+\skips{e'}\geq\skips{e'+1}$. Therefore $e$ is admissible
  for $e'+1$ upon return from the call on $(r'+\skips{e})+\skips{e'}$
  and will be deleted from the list of skip indices. All $e''$ with
  $e>e''>e'$, $\skips{e''}$ will by Observation~\ref{obs:admit}
  likewise be admissible
  ($\skips{e-i}+\skips{e'+i}\leq\skips{e}+\skips{e'}$ for
  $i=0,1,\ldots e-e'$) and removed. Therefore, at the end of the
  \textbf{while}-loop at the call from $r'+\skips{e}$ it will hold
  that $k'=e+1$.
\end{proof}

For the correctness of the receive schedules computed by
Algorithm~\ref{alg:recvschedule}, we define for processor $r$ that
$\sendblock{k}_r = \recvblock{k}_{t^k_r}$ where as in
Algorithm~\ref{alg:broadcast}, $t^k_r=(r+\skips{k})\bmod p$.  With this
definition, the first two correctness conditions from
Section~\ref{sec:broadcasts} are obviously satisfied. It is also clear
from the construction that
$\recvblock{}=\{-1,-2,\ldots,-q\}\setminus\{b-q\}\cup\{b\}$ which
means that all $q$ different blocks can be received over $q$
communication rounds. It needs to be shown that a sent block $\sendblock{k}$
is either the baseblock $b$ or a block that has been received in some earlier
round $j,0\leq j<k$.

\begin{proposition}
  \label{prop:recvcorr}
  The receive schedule blocks computed by Algorithm~\ref{alg:dfspath}
  for any processor $r$ fulfill the correctness condition that
  either $\sendblock{k}=\recvblock{j}$ for some $j, 0\leq j<k$, or
  $\sendblock{k}=b-q$ where $b\geq 0$ is the baseblock for
  processor $r$, $b=\textsc{baseblock(r)}$.
\end{proposition}

\begin{proof}
  We prove that if $\recvblock{k}\neq b$, then
  $\recvblock{k}_r=\recvblock{j}_{r-\skips{k}}$ for some $j,0\leq
  j<k$.
\end{proof}

We summarize the discussion in the following main theorem that states
how to compute the receive schedules for Algorithm~\ref{alg:broadcast}
and Algorithm~\ref{alg:irregallgather}.
\begin{theorem}
  \label{thm:recvsummary}
  A correct receive schedule fulfilling the four correctness
  conditions from Section~\ref{sec:broadcasts} for a $p$-processor
  circulant graph with skips computed by
  Algorithm~\ref{alg:circulants} can be computed in $O(\log p)$ time
  steps for each processor $r,0\leq r<p$.  The receive schedule
  computation from in Algorithm~\ref{alg:recvschedule} can readily be
  implemented.
\end{theorem}

\subsection{The send schedule}
\label{sec:sendschedule}

The straightforward computation of send schedules from the receive
schedules by for processor $r$ setting
$\sendblock{k}_r=\recvblock{k}_{t^k_r}$ with each $\recvblock{k}$
computed by Algorithm~\ref{alg:recvschedule} will take $O(\log^2 p)$
operations. To reach $O(\log p)$ operations, a different approach is
required.  For this structural approach, which will be described in
the following, it is instructive to first consider the case where $p$
is a power of two, $p=2^q$, for which it is well-known how to compute
send (and receive) schedules in $O(q)$ operations~\cite{JohnssonHo89}.

\begin{table}
  \caption{A send schedule for a power-of-two number of processors,
    $p=16$, $q=\log_2 p=4$. The table shows for each processor $r,0\leq
    r<p$ which block to send (to processor $(r+\skips{k})\bmod p$)
    in round $k, k=0,1,2,3$.}
  \label{tab:powertwo}
  \begin{center}
    \begin{tabular}{rrrrrrrrrrrrrrrrr}
      $r$: & 0 & 1 & 2 & 3 & 4 & 5 & 6 & 7 & 8 & 9 & 10 & 11 & 12 & 13 & 14 & 15 \\
    Baseblock $b$ before: & 4 & 0 & 1 & 0 & 2 & 0 & 1 & 0 & 3 & 0 & 1 & 0 & 2 & 0 & 1 & 0 \\
    \toprule
    Sent in round $k=0$: & 4 & 0 & 1 & 0 & 2 & 0 & 1 & 0 & 3 & 0 & 1 & 0 & 2 & 0 & 1 & 0 \\
    Sent in round $k=1$: & 4 & 4 & 1 & 1 & 2 & 2 & 1 & 1 & 3 & 3 & 1 & 1 & 2 & 2 & 2 & 1 \\
    Sent in round $k=2$: & 4 & 4 & 4 & 4 & 2 & 2 & 2 & 2 & 3 & 3 & 3 & 3 & 2 & 2 & 2 & 2 \\
    Sent in round $k=3$: & 4 & 4 & 4 & 4 & 4 & 4 & 4 & 4 & 3 & 3 & 3 & 3 & 3 & 3 & 3 & 3 \\
    \bottomrule
  \end{tabular}
  \end{center}
\end{table}

An example with $p=16$ processors is given in
Table~\ref{tab:powertwo}. We assume that each of the $p$ processors
has already received its baseblock $b, 0\leq b<q$ (with as before
$q=\ceiling{\log_2 p}$). The baseblock $b$ for processor $r,0\leq r<p$
is the largest $b$ such that $r\bmod 2^{b}=0$.  This is also the
baseblock that will be computed by calling $\textsc{baseblock}(r)$
with $\skips{k}=2^k,k=0,1,\ldots \log_2 p$ (as will be computed by
Algorithm~\ref{alg:circulants} for the power-of-two case). The send
schedule will be used to ensure that, after $q$ rounds, each processor
has received all the $q$ different baseblocks. The (unique) send
schedule that ensures this is for processor $r$ to send its own
baseblock to processor $(r+\skips{k})\bmod p$ in rounds $k=0,\ldots,
b$; in rounds $k=b+1,\ldots, q-1$ processor $r$ sends the largest block
received so far. Taking $r$ as a binary number, the block
corresponding to the next set bit in $r\vee p$ (here $\vee$ denotes
bitwise-or) after bit $k-1$ is the block that is sent in round $k$ (in
round $k=0$ the block numbered by the first set (least significant)
bit is sent).

Another way to arrive at the same pattern is to start from round
$k=q-1$ and work downwards to $k=0$.  Let initially $r'=r$, and let
$c$, the block from the previous round, be $q$ (in the example in
Table~\ref{tab:powertwo}, $q=4$). In round $k$, if $r'<\skips{k}$,
send $c$ from the previous round, otherwise ($r'\geq\skips{k}$)
send $c=k$ and let for the next lower round $r'$ be $r'-\skips{k}$.

\begin{algorithm}
  \caption{Computing the send schedule for processors $r=0$ (root) and $r,
    0\leq r<p$.}
  \label{alg:sendschedule}
  \begin{algorithmic}
    \Procedure{Sendschedule}{$r,\sendblock{}$}
    \If{$r=0$}
    \For{$k=0,\ldots,q-1$}\ $\sendblock{k}\gets k$
    \EndFor
    \Else
    \State $b\gets\Call{baseblock}{r}$
    \State $r',c,e \gets r,b,p$
    \For{$k=q-1,\ldots,1$}
    \Comment Obvious invariant: $r'<e$
    \If{$r'<\skips{k}$}
    \State $\ldots$
    \Comment Actions for lower $r'$ here (shown as Algorithm~\ref{alg:lowersend}):
    \If{$e>\skips{k}$} $e\gets \skips{k}$
    \EndIf
    \Else \Comment Here $r'\geq\skips{k}$
    \State $c\gets k-q$
    \State $\ldots$
    \Comment Actions for upper $r'$ here (shown as Algorithm~\ref{alg:uppersend}):
    \State $r',e\gets r'-\skips{k},e-\skips{k}$
    \EndIf
    \EndFor
    \State $\sendblock{0}\gets b-q$
    \EndIf
    \EndProcedure
  \end{algorithmic}
\end{algorithm}

The send schedule computation for arbitrary $p$ (not only
powers-of-two) approximates this behavior (and will be exactly
identical, when $p$ is a power of two). In the power-of-two case, it
will always hold that $0\leq r'<\skips{k+1}$. In round $k$, processor
$r$ sends to processor $(r+\skips{k})\bmod p$, which for
$r'<\skips{k}$ is a processor with $\skips{k}\leq r'<\skips{k+1}$, and
for $r'\geq\skips{k}$ a processor with $r'$ outside this range. The
assumption is that such processors have not received block $c=k$.

The send schedule computation will maintain for processor $r$
when computing its send schedule a virtual processor rank $r'$ and and
upper bound $e$ with $0\leq r'<e$. Starting from round $k=q-1$, $e$ is
initially $\skips{q}=p$ and $r'=r$. In each round, the range of
processors $r'$ is divided into lower part $0\leq r'<\skips{k}$ and
upper part $\skips{k}\leq r'<e$ (which may be empty, if
$e\leq\skips{k}$).  To maintain the invariant for the next round
$k-1$, if $r'$ is in the upper part, both $r'$ and $e$ are decreased
by $\skips{k}$ at the end of round $k$.

The block to be sent in round $k$ is denoted by $c$ and is initially
for $r'$ in the lower part the baseblock $b$ for processor $r$, and for $r'$
in the upper part $c=k$. This outline is shown as
Algorithm~\ref{alg:sendschedule}.

\begin{algorithm}
  \caption{The send schedule computation for iteration $k$ for $r'<\skips{k}$ (lower part).}
  \label{alg:lowersend}
  \begin{algorithmic}
    \If{$e<\skips{k-1}\vee (k=1\wedge b>0)$} $\sendblock{k}\gets c$
    \ElsIf{$r'=0\wedge k=2$}
    \If{$e=2\wedge \skips{2}=3$}
    \State \Call{recvschedule}{$(r+\skips{k})\bmod p,\ablock[]$} \Comment Violation (1)
    \State $\sendblock{k}\gets\ablock[k]$
    \Else\ $\sendblock{k}\gets c$
    \EndIf
    \ElsIf{$r'=0\wedge \skips{k}=5$}  \Comment Implies $k=3$
    \If{$e=3$}
    \State \Call{recvschedule}{$(r+\skips{k})\bmod p,\ablock[]$} \Comment Violation (1)
    \State $\sendblock{k}\gets\ablock[k]$
    \Else\ $\sendblock{k}\gets c$
    \EndIf
    \ElsIf{$r'+\skips{k}\geq e$}
    \State \Call{recvschedule}{$(r+\skips{k})\bmod p,\ablock[]$} \Comment Violation (2)
    \State $\sendblock{k}\gets\ablock[k]$
    \Else\ $\sendblock{k} \gets c$
    \EndIf
    \If{$e>\skips{k}$} $e\gets \skips{k}$
    \EndIf
  \end{algorithmic}
\end{algorithm}

If in round $k$, the $r'$ for processor $r$ is in the lower part,
$r'<\skips{k}$, the processors for which $r'+\skips{k}<e$ have not yet
received block $c$, so $c$ is to be sent if $r'+\skips{k}<e$. Otherwise, it
is not known which block processor $(r+\skips{k})\bmod p$ is missing,
so in that case the receive block for round $k$ for processor
$(r+\skips{k})\bmod p$ is taken as the block to send. This is called a
\emph{violation}, and if there is more than a constant number of such
violations for some processor $r$, a logarithmic number of operations
in total cannot be guaranteed. The lower part is shown as
Algorithm~\ref{alg:lowersend}. The algorithm includes some
observations that can be made for the case where
Observation~\ref{obs:tunnel} holds.  Also, if $e$ is very small,
$e\leq\skips{k-1}$, processor $(r+\skips{k})\bmod p$ will not have
received $c$ and this block can therefore be sent.

\begin{algorithm}
  \caption{The send schedule computation for iteration $k$ for
    $r'\geq\skips{k}$ (upper part).}
  \label{alg:uppersend}
  \begin{algorithmic}
    \If{$k=1\vee r'>\skips{k}\vee e-\skips{k}<\skips{k-1}$} $\sendblock{k}\gets c$
    \ElsIf{$k=2$}
    \If{$\skips{2}=3\wedge e=5$} \Comment Violation (1)
    \State \Call{recvschedule}{$(r+\skips{k})\bmod p,\ablock[]$} 
    \State $\sendblock{k}\gets\ablock[k]$
    \Else\ $\sendblock{k}\gets c$
    \EndIf
    \ElsIf{$\skips{k}=5$} \Comment Implies $k=3$
    \If{$e=8$} \Comment Violation (1)
    \State \Call{recvschedule}{$(r+\skips{k})\bmod p,\ablock[]$}
    \State $\sendblock{k}\gets\ablock[k]$
    \Else\ $\sendblock{k}\gets c$
    \EndIf
    \ElsIf{$r'+\skips{k}>e$} \Comment Violation (3)
    \State \Call{recvschedule}{$(r+\skips{k})\bmod p,\ablock[]$}
    \State $\sendblock{k}\gets\ablock[k]$
    \Else\ $\sendblock{k} \gets c$
    \EndIf
  \end{algorithmic}
\end{algorithm}

If instead $r'$ is in the upper part for round $k$, then only the
processor with $r'=\skips{k}$ may have to use the receive block for
processor $(r+\skips{k})\bmod p$ as the block to send. The upper part
is shown as Algorithm~\ref{alg:uppersend}.

\begin{proposition}
  \label{prop:sendtime}
  Algorithm~\ref{alg:sendschedule} computes for any $r,0\leq r<p$ a
  send schedule in $O(\log p)$ operations.
\end{proposition}
\begin{proof}
  The loop performs $q-1$ iterations.  Iterations that are not
  violations take constant time. We will show that there are only a
  constant number of violations of the form (1-3), namely at most four
  ($4$). Each violation takes $O(\log p)$ steps by the receive schedule
  Proposition~\ref{prop:recvtime}. Therefore, the send schedule
  computation takes $\Theta(\log p)$ steps.

  All violations (1) and (3) for the upper part case where
  $r'\geq\skips{k}$ for some iteration $k$ can happen at most once,
  since for all such possible violations it holds that $r'=\skips{k}$ by
  the first condition (indeed, $\sendblock{k}=c$ when
  $r'>\skips{k}$). After the end of an iteration where such a
  violation (1) or (3) could have happened, $r'=0$ for all remaining
  iterations, thus it will never again hold that $r'\geq\skips{k}$.

  We therefore only have to consider violations (1) and (2) for the
  lower part case where $r'\leq\skips{k}$.  Violations (1) can happen
  only in the two iterations $k=2$ and $k=3$ (and here only for
  $r'=0$). Violation (2) can happen for $k=1$, and $k>2$.  This
  violation happens if $r'+\skips{k}\geq e$. If $r'<\skips{k-1}$ this
  violation can possibly happen again at iteration $k-1$, if also
  $r'+\skips{k-1}\geq e'$ where $e'$ is the upper bound for iteration
  $k-1$. However, for $e>\skips{k}$, the upper bound $e'$ is
  $\skips{k}$, so $r'+\skips{k-1}>\skips{k}$ can happen only for
  $r'=\skips{k-1}$ (which is then taken care of in the upper part for
  iteration $k-1$). Therefore, only the cases where $e\leq\skips{k}$
  have to be considered. By repeated use of
  Observation~\ref{obs:skips}, it is easy to see that the smallest
  possible $e$ for iteration $k$ is
  $\skips{k+1}-(q-1-k)=\skips{k+1}-(q-1)+k$. A violation (2) can
  happen for $\skips{k-1}\leq e\leq\skips{k}$. For the smallest
  possible $e$, we have $\skips{k+1}-(q-1)+k\leq\skips{k}$ implying
  $\skips{k+1}-\skips{k}\leq (q-1)+k$ and with
  Observation~\ref{obs:skips} $\skips{k}\leq q-k$ which implies also
  $2\skips{k-1}-2\leq (q-1)-k$. Likewise,
  $\skips{k-1}\leq e$ gives $\skips{k-1}\leq\skips{k+1}-(q-1)+k$
  implying $(q-1)-k\leq\skips{k+1}-\skips{k-1}$. Using again
  Observation~\ref{obs:skips}, we can simplify this into
  $(q-1)-k\leq 3\skips{k-1}$.  The inequalities are fulfilled for at most one
  iteration $k$.
  
  The possibility that a violation of type (2) happens in the iteration $k-2$
  ($r'$ is in the lower part, then in the upper part, then in the lower part
  again) can be excluded. In iteration $k-2$ it would then have to hold that
  $r'-\skips{k-1}\geq e-\skips{k-1}$ which is per the invariant that $r'<e$
  not possible.
\end{proof}

\begin{table}
  \caption{Receive and send schedule for a non-power-of-two number of
    processors, $p=17$, $q=\ceiling{\log_2 p}=5$.  The table shows for
    each processor $r,0\leq r<p$ the baseblock $b$ and the $\recvblock{k}$ and
    $\sendblock{k}$ schedules for $k=0,1,2,3,4$.}
  \label{tab:p17}
  \begin{center}
  \begin{tabular}{rrrrrrrrrrrrrrrrrr}
$r$: &  0 & 1 & 2 & 3 & 4 & 5 & 6 & 7 & 8 & 9 & 10 & 11 & 12 & 13 & 14 & 15 & 16 \\
$b$: &5 & 0 & 1 & 2 & 0 & 3 & 0 & 1 & 2 & 4 & 0 & 1 & 2 & 0 & 3 & 0 & 1 \\
\toprule
$\recvblock{0}$: & -4 & 0 & -5 & -4 & -3 & -5 & -2 & -5 & -4 & -3 & -1 & -5 & -4 & -3 & -5 & -2 & -5 \\
$\recvblock{1}$: & -5 & -4 & 1 & -5 & -4 & -3 & -3 & -2 & -5 & -4 & -3 & -1 & -5 & -4 & -3 & -3 & -2 \\
$\recvblock{2}$: & -2 & -2 & -2 & 2 & 0 & -4 & -4 & -3 & -2 & -2 & -4 & -3 & -1 & -1 & -4 & -4 & -3 \\
$\recvblock{3}$: & -1 & -3 & -3 & -2 & -2 & 3 & 0 & 1 & 2 & -5 & -2 & -2 & -2 & -2 & -1 & -1 & -1 \\
$\recvblock{4}$: & -3 & -1 & -1 & -1 & -1 & -1 & -1 & -1 & -1 & 4 & 0 & 1 & 2 & 0 & 3 & 0 & 1 \\
\midrule
$\sendblock{0}$: & 0 & -5 & -4 & -3 & -5 & -2 & -5 & -4 & -3 & -1 & -5 & -4 & -3 & -5 & -2 & -5 & -4 \\
$\sendblock{1}$: & 1 & -5 & -4 & -3 & -3 & -2 & -5 & -4 & -3 & -1 & -5 & -4 & -3 & -3 & -2 & -5 & -4 \\
$\sendblock{2}$: & 2 & 0 & -4 & -4 & -3 & -2 & -2 & -4 & -3 & -1 & -1 & -4 & -4 & -3 & -2 & -2 & -2 \\
$\sendblock{3}$: & 3 & 0 & 1 & 2 & -5 & -2 & -2 & -2 & -2 & -1 & -1 & -1 & -1 & -3 & -3 & -2 & -2 \\
$\sendblock{4}$: & 4 & 0 & 1 & 2 & 0 & 3 & 0 & 1 & -3 & -1 & -1 & -1 & -1 & -1 & -1 & -1 & -1 \\
\bottomrule
  \end{tabular}
\end{center}
\end{table}

Finite, exhaustive proof for $p$ up to some millions shows that the number
of violations is indeed at most $4$ (but sometimes $3$), see the
discussion in Section~\ref{sec:empirics}. Table~\ref{tab:p17} shows
receive and send schedules as computed by the algorithms for $p=17$
processors (not a power of two). There are, for instance, send
schedule violations in the sense of Algorithm~\ref{alg:lowersend} in
round $k=2$ for processor $r=3$ and in round $k=3$ for processor
$r=8$.

\begin{proposition}
  \label{prop:sendcorr}
  The send schedule computed by Algorithm~\ref{alg:sendschedule} is correct.
\end{proposition}

\begin{proof}
  It will be shown that $\sendblock{k}_r =
  \recvblock{k}_{r+\skips{k}}$ for $k=0,1,\ldots,q-1$.  The first
  block $\sendblock{0}=b-q$ is obviously correct. For the cases where
  there is a violation, $\sendblock{k}_r$ is computed as
  $\recvblock{k}_{r+\skips{k}}$, so also these cases are obviously
  correct.
\end{proof}

\section{Empirical Results}
\label{sec:empirics}

\begin{table}
  \caption{Timings of old $O(\log^3 p)$ and new $O(\log p)$ time step
    receive and send schedule algorithms for different ranges of
    processors $p$. Receive and send schedules are computed for all
    processors $0\leq r<p$ for all $p$ in the given ranges. The
    expected running times are thus bounded by $O(p\log^3 p)$ (old)
    and $O(p \log p)$ (new) time, respectively.  Times are in seconds
    and measured with the \texttt{clock()} function. We also estimate
    the average time spent per processor for computing its send and
    receive schedules of $\ceiling{\log_2 p}$ entries. This is done by
    measuring for each $p$ the total time for the schedule
    computation, dividing by $p$ and averaging over all $p$ in the
    range. These times are in micro seconds.}
  \label{tab:runtimes}
  \begin{center}
    \begin{tabular}{crrrr}
      Range of processors $p$ &
      \multicolumn{2}{c}{Total Time (seconds)} &
      \multicolumn{2}{c}{Per processor ($\mu$seconds)} \\
      & $O(p\log^3 p)$ & $O(p\log p)$ & $O(\log^3 p)$ & $O(\log p)$ \\
      \toprule
      $[1,17\,000]$ & 443.8 & 50.0 & 2.769 & 0.334 \\
      $[16\,000,33\,000]$ & 1567.2 & 152.8 & 3.763 & 0.370 \\
      $[64,000\,73\,000]$ & 3206.0 & 282.6 & 5.187 & 0.454 \\
      $[131\,000,140\,000]$ & 7595.0 & 653.2 & 6.226 & 0.534 \\
      $[262\,000,267\,000]$ & 9579.4 & 726.6 & 7.242 & 0.548 \\
      $[524\,000,529\,000]$ & 21580.2 & 1492.9 & 8.196 & 0.566 \\
      $[1\,048\,000,1\,050\,000]$ & 18934.3 & 1083.8 & 9.024 & 0.516 \\
      $[2\,097\,000,2\,099\,000]$ & 44714.9.0 & 2554.6 & 10.656 & 0.608 \\
      \bottomrule
    \end{tabular}
    \end{center}
\end{table}

The algorithms for computing receive and send schedules in $O(\log p)$
time steps have been implemented for use in implementations for
\mpibcast and \mpiallgatherv. All implementations are available from
the author. To first demonstrate the practical impact of the
improvement from $O(\log^3 p)$ time steps (per processor) which was
the bound given in~\cite{Traff22:bcast,Traff22:bcastinprogress} to
$O(\log p)$ steps per processor as shown here, we run the two
algorithms for different ranges of processors $p$. For each $p$ in
range, we compute both receive and send schedules for all processors
$r, 0\leq r<p$, and thus expect total running times bounded by
$O(p\log^3 p)$ and $O(p\log p)$, respectively.  These runtimes in
seconds, gathered on a standard workstation with an Intel Xeon E3-1225
CPU at 3.3GHz and measured with the \texttt{clock()} function from the
\texttt{time.h} C library, are shown in Table~\ref{tab:runtimes}. The
timings include both the receive and the send schedule computations,
but exclude the time for verifying the correctness of the schedules,
which has also been performed up to some $p\geq 2M, M=2^{20}$ (and for
a range of $100\,000$ processors around $16M$ which took about a
week), including verifying the bounds on the number of recursive calls
from Proposition~\ref{prop:recvtime} and the number of violations from
Proposition~\ref{prop:sendtime}. When receive and send schedules have
been computed for all processors $r,0\leq r<p$, verifying the four
correctness conditions from Section~\ref{sec:broadcasts} can obviously
be done in $O(p\log p)$ time steps. The difference between the old and
the new implementations is significant, from close to a factor of 10
to significantly more than a factor of 10. However, the difference is
not by a factor of $\log^2 p$ as would be expected from the derived
upper bounds.  This is explained by the fact that the old send
schedule implementations employ some improvements beyond the trivial
computation from the receive schedules which makes the complexity
closer to $O(\log^2 p)$. These improvements were not documented
in~\cite{Traff22:bcast,Traff22:bcastinprogress}, but can be found in
the actual code. The old receive schedule computation
from~\cite{Traff22:bcast,Traff22:bcastinprogress} is in $O(\log^2 p)$,
though. We also give a coarse estimate of the time spent per processor
by measuring for each $p$ in the given range the time for the schedule
computations for all $p$ processors, dividing this by $p$ and
averaging over all $p$ in the given range. This is indicative of the
overhead for the \textsc{recvblock()} and \textsc{sendblock()}
computations in the implementations of Algorithm~\ref{alg:broadcast}
and Algorithm~\ref{alg:irregallgather}. These times (in microseconds)
are listed as columns $O(\log^3 p)$ and $O(\log p)$, respectively.
The difference by a factor of $10$ and more is slowly increasing with
$\log_2 p$.

Preliminary experiments with \mpibcast and \mpiallgatherv
implementations following closely Algorithm~\ref{alg:broadcast} and
Algorithm~\ref{alg:irregallgather} were given previously
in~\cite{Traff22:bcast,Traff22:bcastinprogress} as well, and we for
completeness run the same kind of experiments with the new schedule
computations. Our system is a small $36\times 32$ processor cluster
with 36~dual socket compute nodes, each with two Intel(R) Xeon(R) Gold
6130F 16-core processors. The nodes are interconnected via dual Intel
Omnipath interconnects each with a bandwidth of 100 GigaBytes/s. The
implementations and benchmarks were compiled with \gcc with the
\texttt{-O3} option.

The best number of blocks $n$ leading to the smallest broadcast time
is chosen based on a linear cost model. For \mpibcast, the size of the
blocks is chosen as $F\sqrt{m/\ceiling{\log p}}$ for a constant $F$
chosen experimentally. For \mpiallgatherv, the number of blocks to be
used is chosen as $\sqrt{m\ceiling{\log p}}/G$ for another,
experimentally determined constant $G$.  The constants $F$ and $G$ depend on
context, system, and MPI library. Finding a best $n$ in practice
is a highly interesting problem outside the scope of this work.
Likewise, using the implementations on clustered, hierarchical
systems, for instance as suggested in~\cite{Traff20:mpidecomp}, is
likewise open and will be dealt with elsewhere.

\begin{figure*}
  \begin{center}
    \includegraphics[width=.3\linewidth]{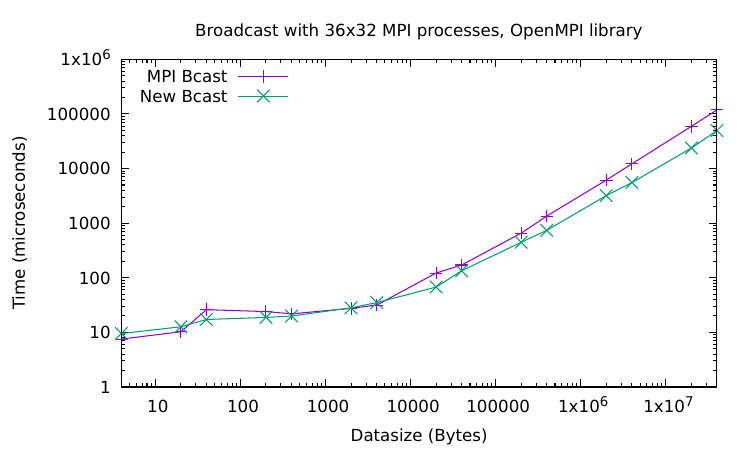}
    \includegraphics[width=.3\linewidth]{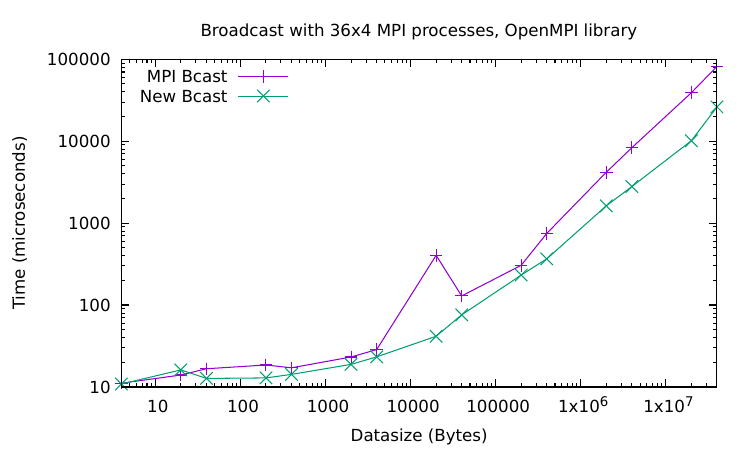}
    \includegraphics[width=.3\linewidth]{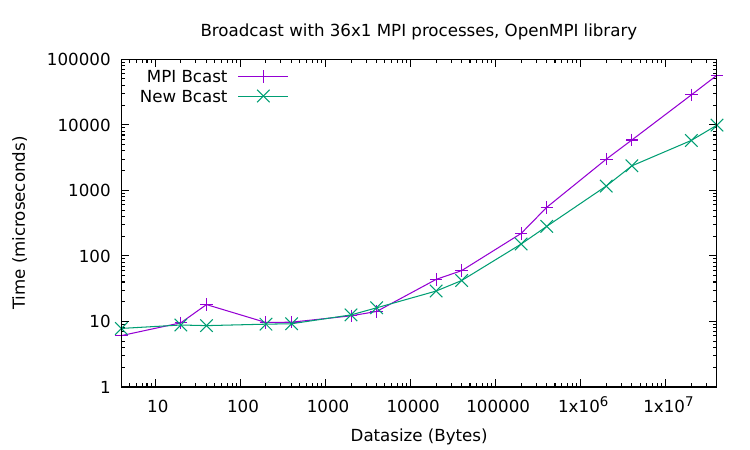}
  \end{center}
  \caption{Broadcast results, native versus new, with the
    \hydraopenmpi library with $p=36\times 32, p=36\times 4, p=36\times 1$
    MPI processes.  The constant factor $F$ for the size of the blocks
    has been chosen as $F=70$. The MPI datatype is \mpiint.}
  \label{fig:bcastopenmpi}
\end{figure*}

\begin{figure*}
  \begin{center}
    \includegraphics[width=.3\linewidth]{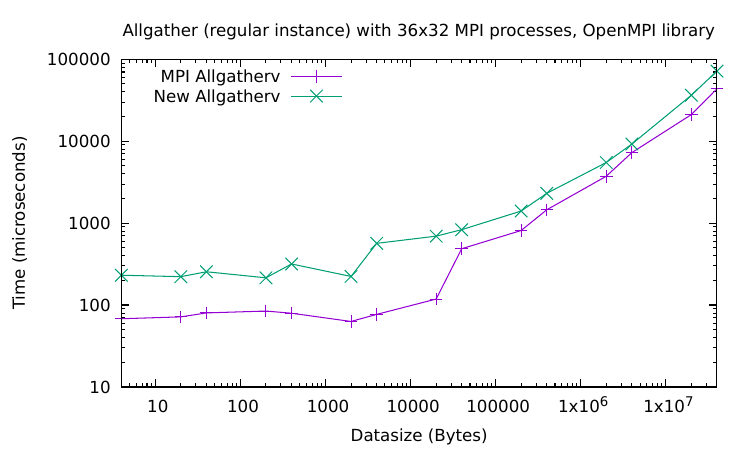}
    \includegraphics[width=.3\linewidth]{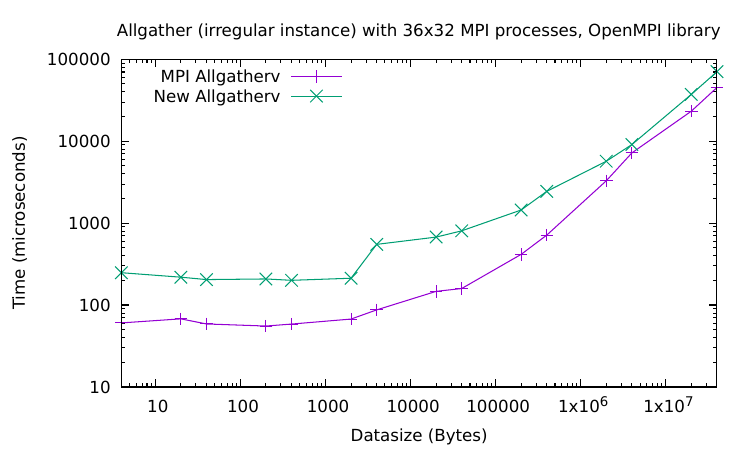}
    \includegraphics[width=.3\linewidth]{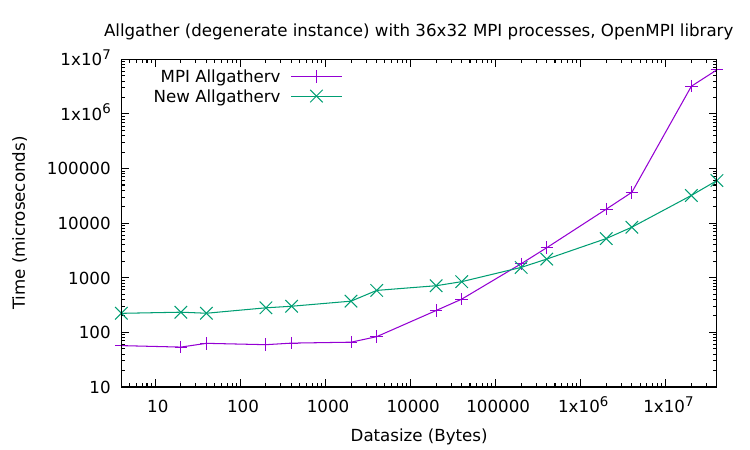}
  \end{center}
  \caption{Irregular allgather results, native versus new, with the
    \hydraopenmpi library with $p=36\times 32$ MPI processes and
    different types of input problems (regular, irregular,
    degenerate).  The constant factor $G$ for the number of blocks has
    been chosen as $G=40$. The MPI datatype is \mpiint.}
  \label{fig:allgatopenmpi}
\end{figure*}

\begin{figure*}
  \begin{center}
    \includegraphics[width=.3\linewidth]{allgatreg36x32openmpi.pdf}
    \includegraphics[width=.3\linewidth]{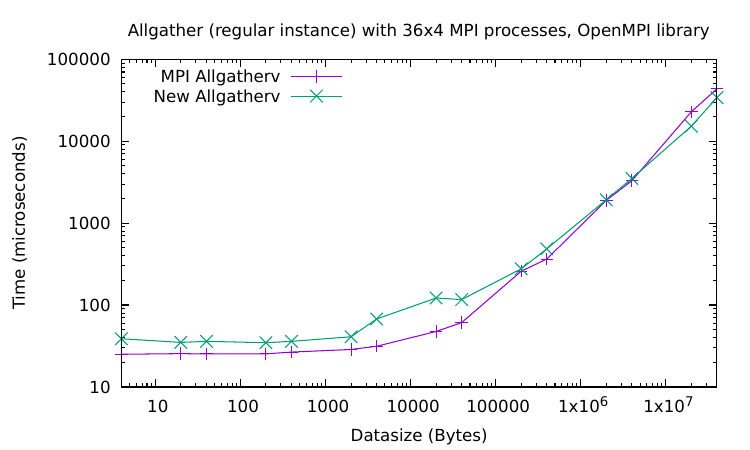}
    \includegraphics[width=.3\linewidth]{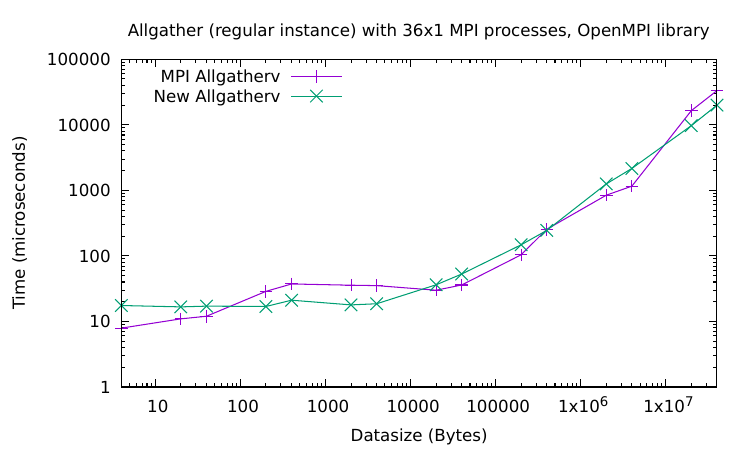}
  \end{center}
  \caption{Regular allgather results, native versus new, with the
    \hydraopenmpi library with $p=36\times 32, p=36\times 4, p=36\times 1$
    MPI processes. The constant factor $G$ for the number of blocks
    has been chosen as $G=40$. The MPI datatype is \mpiint.}
  \label{fig:allgatregopenmpi}
\end{figure*}

The results for \mpibcast are shown in Figure~\ref{fig:bcastopenmpi}
for the full $36$ nodes of the system and different number of MPI
processes per node, namely $32,4$ and $1$. It is noteworthy that the new
implementation which assumes homogeneous communication can be
significantly faster than the \hydraopenmpi baseline library
implementation even for the $36\times 32$ process case.

The results for \mpiallgatherv are shown in
Figure~\ref{fig:allgatopenmpi} for $p=36\times 32$ MPI processes and
different types of input problems. The \emph{regular problem} divides
the given input size $m$ roughly evenly over the processes in chunks
of $m/p$ elements. The \emph{irregular problem} divides the input in
chunks of size roughly $(i\bmod 3) m/p$ for process $i=0,1,\ldots,
p-1$. The \emph{degenerate problem} has one process contribute the
full input of size $m$ and all other process no input elements. For
the degenerate problem, the performance of the \hydraopenmpi baseline
library indeed degenerates and is a factor of close to 100 slower than the
new implementation, where the running time is largely independent of
the problem type. The running time of the new \mpiallgatherv
implementation is in the ballpark of \mpibcast for the same total
problem size.  For completeness, Figure~\ref{fig:allgatregopenmpi}
gives the running times for the regular problems with fewer MPI
processes per node, $p=36\times 4$ and $p=36\times 1$.

\section{Summary}

We showed that round-optimal broadcast schedules on fully connected,
one-ported, fully bidirectional $p$-processor systems can indeed be
computed in $O(\log p)$ time steps per processor. This affirmatively
answers the long standing open questions posed
in~\cite{Traff08:optibcast,Traff22:bcastinprogress,Traff22:bcastba,Traff22:bcast}. We
repeated experiments indicating that the computations are feasible for
use in practical implementations of \mpibcast and \mpiallgatherv. A
more careful evaluation of these implementations, also in versions
that are more suitable to systems with hierarchical, non-homogeneous
communication systems is ongoing and should be found elsewhere.

For the full $O(\log p)$-sized schedule computations an overhead of
$O(\log p)$ is incurred, but complexity per neighboring processor is
only $O(1)$ amortized. Would it be possible to find each send and
receive block in $O(1)$ worst-case time? Also interesting is to
characterize when the schedules are unique, how many different
schedules there are for a given $p$, and for which
$\ceiling{\log_2 p}$-regular circulant graphs the constructions can work.

\bibliographystyle{plain}
\bibliography{traff,parallel} 

\end{document}